\documentclass[a4paper,12pt,english]{article}
\usepackage{amsfonts}
\usepackage{graphicx}
\usepackage{amsmath}
\usepackage{color}

\newcommand{\be}{\begin{equation}}
\newcommand{\ee}{\end{equation}}
\begin{document}


\begin{titlepage}
\begin{center}

\noindent{{\LARGE{The symmetries of magnetized horizons}}}

\smallskip
\smallskip

\smallskip
\smallskip
\smallskip
\smallskip
\smallskip
\smallskip
\noindent{\large{Sasha Brenner, Gaston Giribet, Luciano Montecchio}}

\smallskip
\smallskip

\smallskip
\smallskip

\smallskip
\smallskip
\centerline{Physics Department, University of Buenos Aires FCEyN-UBA and IFIBA-CONICET}
\centerline{{\it Ciudad Universitaria, pabell\'on 1, 1428, Buenos Aires, Argentina.}}

\end{center}

\bigskip

\bigskip

\bigskip

\bigskip

\begin{abstract}
We study stationary black holes in the presence of an external strong magnetic field. In the case where the gravitational backreaction of the magnetic field is taken into account, such an scenario is well described by the Ernst-Wild solution to Einstein-Maxwell field equations, representing a charged, stationary black hole immersed in a Melvin magnetic universe. This solution, however, describes a physical situation only in the region close to the black hole. This is due to the following two reasons: Firstly, Melvin spacetime is not asymptotically locally flat; secondly, the non-static Ernst-Wild solution is not even asymptotically Melvin due to the infinite extension of its ergoregion. All this might seem to be an obstruction to address an scenario like this; for instance, it seems to be an obstruction to compute conserved charges as this usually requires a clear notion of asymptotia. Here, we circumvent this obstruction by providing a method to compute the conserved charges of such a black hole by restricting the analysis to the near horizon region. We compute the Wald entropy, the mass, the electric charge, and the angular momentum of stationary black holes in highly magnetized environments from the horizon perspective, finding results in complete agreement with other formalisms.   
\end{abstract}
\end{titlepage}


\section{Introduction}

Compact objects in strong magnetic fields are of great importance in astrophysics. Scenarios such as supermasive black holes in active galactic nuclei, or even at the center of our galaxy \cite{MW} could be of that sort. One can also think of other scenarios, such as binary systems involving a stellar black hole and a magnetar \cite{Magnetars, Magnetars2}. In the latter case, for example, the energy density of the magnetic field in close proximity to the star would be enormous, exceeding in many order of magnitudes the mass density of heavy elements, and so yielding a non-negligible gravitational backreaction. Here, we will study the case of black holes placed in a region of very strong magnetic fields, including the gravitational backreaction of the latter on the spacetime geometry. Such a highly magnetized region of the spacetime may well be approximated by the Melvin solution to Einstein equations \cite{Melvin}, an axisymmetric geometry that describes a universe filled with a stable bundle of magnetic lines bounded by gravitational interaction \cite{Melvin2, Thorne}. In a Melvin type spacetime one may consider to place a black hole and to use such scenario to model how the magnetic field affects the near horizon region. The solution describing a stationary black hole in a Melvin universe is known analytically; it has been given by Ernst and Wild in \cite{Ernst, ErnstWild}; see also \cite{Wald, Hiscock}. The solution, however, describes a physical situation as long as one focuses on the region close to the black hole. This is due to the non-standard asymptotics of the Melvin spacetime, which is not asymptotically flat; in addition, as shown in \cite{Gibbons}, Ernst-Wild solution is not strictly asymptotically Melvin due to the infinite extension of its ergoregion. All this might seem to represent an obstruction to address an scenario like this, as in gravitational theories one frequently deals with computations that demand to have a clear notion of asymptotia, an example being the standard methods to compute conserved charges. This is precisely why the possibility of accomplishing a robust method to compute conserved charges from the near horizon perspective becomes particularly interesting \cite{DGGP1}. Here, by extending the near horizon symmetry methods developed in \cite{DGGP1, DGGP2, Puhm} to the case of magnetized horizons\footnote{The near horizon limit of extremal and near-extremal black holes in Melvin universe has been considered in the literature; see for instance \cite{Astorino, Astorino2, Bicak1} and references therein and thereof. Here, in contrast, we will not restrict the analysis to the set of extremal configurations.}, we will be able to explicitly derive the relevant charges that describe the physics of charged, stationary black holes in strong magnetic fields. The Wald entropy, the angular momentum, and electric charges of the Ernst-Wild black hole solution in Melvin universe will be obtained in a direct way. To do so, we will first show that such a solution can be accommodated in the set of asymptotic near horizon conditions introduced in \cite{Puhm} for the Einstein-Maxwell theory, so yielding an infinite-dimensional symmetry algebra in the vicinity of magnetized horizons. Then, the black hole charges will be shown to coincide with the Noether charges associated to such symmetry\footnote{Conserved charges and thermodynamics of black holes in Melvin universes have been studied with other methods; for instance, in references \cite{Hawking, Gibbons, Gibbons2, AstorinoCompere}. Our formalism, while qualitatively different, leads to results that are consistent with those of previous analysis}. 

The paper is organized as follows: In section 2, we will review the Ernst solution to Einstein equations \cite{Ernst}, which describe a black hole in a backreacting external magnetic field. We will focus on the physical interpretation of the spacetime geometry and comment the most important aspects of the geometry. In section 3, as a first example, we will show how such a solution can be studied with the near horizon techniques developed in \cite{DGGP1}. This will lead us to generalize the results previously obtained in \cite{DGGP2} and \cite{Puhm} to the case where the conserved charges depend on the backreacting external magnetic field. In section 4, we will extend the analysis to the case of electrically charged, stationary black hole solutions immersed in a strong external magnetic field. This shows that also this case admits a near horizon description as the one in \cite{DGGP1}, yielding supertranslations and superrotation charges. In section 5, we will show that, while the zero-modes of the supertranslation charges are shown to reproduce the Wald entropy of the black holes, the superrotation charges at the horizon will correctly reproduce the angular momentum contribution of the electromagnetic field, manifestly exhibiting a Gauss phenomenon. Our computation of the Wald entropy, the mass, the angular momentum, and electric charge of the magnetized black hole is found to be in complete agreement with the results obtained by other methods and thus consistent with black hole thermodynamics. In section 6, we briefly discuss the case of magnetically charged black holes in an external field. 

\section{Black holes in a magnetized environment}

Let us start by reviewing the Melvin solution to Einstein-Maxwell equations, which describes an axisymmetric universe filled with a magnetic field \cite{Melvin}. The metric of such spacetime in polar coordinates is given by
\begin{equation}\label{Melvin}
    ds^2=\ \Lambda^2(r,\theta )(\, -dt^2+dr^2+r^2 d\theta^2\, )+\Lambda^{-2}(r,\theta )\, r^2 \sin^2{\theta} d\phi^2 
\end{equation}
with
\begin{equation}\label{Melvin2}
\Lambda(r,\theta )= 1+\frac{1}{4}B_0^2 r^2 \sin^2{\theta},
\end{equation}
where $t\in \mathbb{R}$, $r\in \mathbb{R}_{\geq 0}$, $\phi \in [0,2\pi ]$, $\theta =[0, \pi ]$. It may be useful to consider cylindrical coordinates $\rho = r\, \sin\theta$, $z=r\, \cos \theta $. The magnetic field is given by
\begin{equation}\label{magnetic}
    H_z=\ \Lambda^{-2}(r,\theta )\, B_0\, ,
\end{equation}
with $H_{\rho} = H_{\phi}=H_{\theta} =0$. $B_0$ is a constant that controls the magnitude of the external magnetic field. This solution can be thought of as a bundle of magnetic lines in the $z$-direction that are bounded by gravitational interaction, forming in such a way a stable configuration \cite{Melvin2, Thorne}. The resulting spacetime exhibits isometry group $ISO(1,1)\times SO(2)$, which correspond to the Poincar\'e group in the (1+1)-dimensional constant-$\phi $, constant-$\rho $ slices, together with rotations in $\phi $. While the solution with $B_0=0$ corresponds to Minkowski spacetime, when $B_0\neq 0$ the geometry is not asymptotically flat; in fact, the squashing of the constant-$t$, constant-$r$ surfaces produced by the function $\Lambda(r,\theta)$ increases at large distance, while it becomes negligible when $r\ll B_0^{-1}$. 

By using a solution generating technique developed by Harrison in \cite{Harrison}, Ernst found in \cite{Ernst} a generalization to the Melvin solution that come to describe a static black hole immersed in an external magnetic field, cf. \cite{Wald, Hiscock}. The metric of the solution takes the form
\begin{equation}\label{Ernst}
    ds^2=\ \Lambda^2(r,\theta )\, \Big(\,-f(r)dt^2+\frac{dr^2}{f(r)}+r^2 d\theta^2\,\Big) +\Lambda^{-2}(r,\theta )\,r^2 \sin^2{\theta} d\phi^2
\end{equation}
with
\begin{equation}
\Lambda  (r,\theta )=\ 1+\frac{1}{4}B_0^2 r^2 \sin^2{\theta}\, , \ \ \text{and}\ \ f(r)=\ 1-\frac{2M}{r},
\end{equation}
while the non-vanishing components of the magnetic field are given by
\begin{align}\label{Ernst2}
    H_r =&\ \Lambda^{-2}(r,\theta )\, B_0 \cos{\theta}\\
    H_\theta =&\ -\Lambda^{-2}(r,\theta )\, B_0\Big{(}1-\frac{2M}{r}\Big{)}^{1/2}\sin{\theta}.
\end{align}

It is clear from these expressions that Ernst solution becomes Melvin universe (\ref{Melvin})-(\ref{magnetic}) when $M=0$, while it reduces to the metric of a Schwarzschild black hole of mass $M$ when the external magnetic field vanishes, $B_0=0$ (here we use natural units $G=c=1$). The geometry with $B_0\neq 0\neq M$ exhibits isometry group $\mathbb{R}\times SO(2)$ generated by the Killing vectors $\partial_t$, $\partial_{\phi}$. The interpretation of the metric is that of a Schwarzschild black hole in an asymptotically Melvin universe. The location of the horizon of the black hole is $r_+=2M$, where the spacetime is regular. The geometry is singular only at $r=0$, provided $M\neq 0$. 

Ernst solution (\ref{Ernst})-(\ref{Ernst2}) admits a stationary, charged generalization \cite{Ernst, ErnstWild}, which is the one on which we will focus in this work. It is worth mentioning already here that the electrically charged solution, even when no intrinsic rotation is present, happens to be non-static, but stationary. This is due to the non-vanishing Poynting density of a charged particle in an external magnetic field, which produces an effective dragging in spacetime. This is related to what happens when approaching a magnetic monopole to a Reissner-Nordstr\"om black hole, cf. \cite{HT}. We will discuss the electrically charged black hole case in section 4. Before that, in the next section, we will analyze the symmetries emerging in the near horizon limit of magnetized black holes (\ref{Ernst})-(\ref{Ernst2}). 

\section{Near horizon asymptotics}

To analyze the near horizon symmetries and the associated conserved charges in Einstein-Maxwell theory, we resort to the results of reference \cite{Puhm}, which come to generalize the near horizon analysis of \cite{DGGP1, DGGP2} to the case of electrically, magnetically charged solutions\footnote{Infinite-dimensional near horizon symmetries have also been studied in reference \cite{Hawking:2016msc, Hawking:2016sgy, Afshar:2016uax, Afshar:2017okz, Grumiller:2018scv, Grumiller:2019fmp, Grumiller:2019ygj, Grumiller:2020vvv, Yo}.}. In \cite{Puhm}, the following asymptotic conditions for the metric near the horizon were considered
\begin{align}
    g_{vv}=&\ -2\kappa\, \rho + \mathcal{O}(\rho^2),\nonumber \\
    g_{vA}=&\ \theta_A(z^B)\, \rho+\mathcal{O}(\rho^2),\label{Estas}\\
    g_{AB}=&\ \Omega_{AB}(z^C)+\lambda_{AB}(z^C)\, \rho +\mathcal{O}(\rho^2),\nonumber
\end{align}
together with the gauge fixing condition
\begin{equation}
    g_{\rho\rho}=0,\ \ \ g_{v\rho}=1,\ \ \ g_{A\rho}=0.\label{gaugefixing}
\end{equation}
The horizon location is $\rho =0$, so that $\rho \in \mathbb{R}_{\geq 0}$ is the coordinate that measures the distance from the horizon. Power expansion in $\rho $ thus controls the near horizon asymptotic conditions, with $\mathcal{O}(\rho^n)$ standing for orders that damp off as fast as $\sim \rho^n$, or faster, as one approaches the horizon. The spacetime metric is $ds^2=g_{\mu \nu}dx^{\mu}dx^{\nu}$ ($\mu,\nu=0,1,2,3$), with $x^0=v$, $x^A=z^A$ ($A,B,C=1,2$), and $x^3=\rho $. Coordinate $v\in \mathbb{R}$ represents the null direction at the horizon $\mathcal{H}^{+}=\Sigma_2\times \mathbb{R}$, while $z^A$ ($A=1,2$) represent the angular variables on the constant-$v$ surface $\Sigma_2$. In (\ref{Estas})-(\ref{gaugefixing}), functions $\kappa$, $\theta_A$, $\Omega_{AB}$, $\lambda_{AB}$ may in principle depend on $v$ and on $z^A$. However, as $\kappa$ represents the surface gravity, it turns out to be consant for isolated horizons. Integrability of the Noether charges, on the other hand, demands $\theta_A$ and $\Omega_{AB}$ not to depend on $v$; see \cite{DGGP2} for details. 

The near horizon expansion of the electromagnetic field $A=A_{\mu}dx^{\mu}$, on the other hand, is given by  
\begin{align}
    &A_{v} = A_{v}^{(0)} +   A_{v}^{(1)}(v,z^{A})\, \rho + \mathcal{O}(\rho^2),\label{Estas3}\\
    &A_{B} = A_{B}^{(0)}(z^{A}) +  A_{B}^{(1)}(v,z^{A}) \, \rho + \mathcal{O}(\rho^2),\nonumber
\end{align}
together with the gauge fixing condition $A_\rho=0$. 

Now, let us consider the group of diffeomorphisms and gauge transformations that preserve the above near horizon form for the fields $g_{\mu\nu}$ and $A_{\mu}$. These are generated by the Killing vectors $\xi =\xi^{\mu}\partial_{\mu}$ and gauge functions $\epsilon$ such that, after performing the changes $g_{\mu\nu}\to g_{\mu\nu}+\delta g_{\mu\nu}=g_{\mu\nu}+\mathcal{L}_{\xi} g_{\mu\nu}$, $A_{\mu}\to A_{\mu}+\delta A_{\mu}= A_{\mu }+ \mathcal{L}_{\xi}A_{\mu }+ \partial_{\mu}\epsilon $, with $\mathcal{L}_{\xi}$ being the Lie derivative with respect to $\xi $, leave the expansion in powers of $\rho $ described above unchanged, regardless whether or not the specific functions $\theta_A$, $\Omega_{AB}$, $A_{v}^{(i)}$, $A_{B}^{(i)}$ have changed. The explicit form of $\xi^{\mu} $ and $\epsilon $ that satisfy this can be found in \cite{Puhm}, and it comprehends the expansion 
\begin{eqnarray}
\xi ^v = T(z^A)+\mathcal{O}(\rho), \ \ \ \xi ^{\rho} = \mathcal{O}(\rho), \ \ \ \xi ^A = Y^A(z^B)+\mathcal{O}(\rho),
\end{eqnarray}
and 
\begin{eqnarray}
\epsilon =U(z^A)-T(z^A)A_{v}^{(0)}+\mathcal{O}(\rho), 
\end{eqnarray}
where $T(z^A)$, $Y^A(z^B)$ and $U(z^A)$ are four arbitrary functions of the angular coordinates $z^A$ (say $z^1=z^1( \theta, \phi )$ and $z^2=z^2( \theta, \phi )$). These functions can be expanded in Fourier modes as follows
\begin{align}
    T(z,\bar{z})=&\ \sum_{m,n}T_{(m,n)}\, z^m\bar{z}^n,\nonumber\\
    Y^z(z)=&\ \sum_n{ Y_n\, z^n}\, ,\ \ \ Y^{\bar{z}}(\bar{z})=\sum_n{\bar{Y}_n}\, \bar{z}^n ,\label{Expansion}\\
    U(z,\bar{z})=&\ \sum_{m,n}{U_{(m,n)}\, z^m\bar{z}^n}\, , \nonumber
\end{align}
where $z$ and $\bar{z}$ are now complex variables, related to the angular coordinates $\theta$ and $\varphi$ in a way that depends on the solution under consideration. In (\ref{Expansion}), $m,n\in\mathbb{Z}$, and $T_{(m,n)}$, $Y_n$, $\bar{Y}_n$ and $U_{(m,n)}$ can be regarded as Fourier modes. As the diffeomorphisms and gauge transformations generated by $\xi^\mu$ and $\epsilon$ involve arbitrary functions of the angles, they can be shown to expand an infinite-dimensional algebra, which consists of two copies of Witt algebra in semi-direct sum with two sets of Abelian currents. More precisely, $Y_n$ and $\bar{Y}_n$ generate Witt algebra and so they are called superrotations; while the two Abelian currents are generated by $T_{(m,n)}$ and $U_{(m,n)}$, and they are called supertranslations; see \cite{DGGP1, DGGP2, Puhm, Yo}. 

The infinite-dimensional symmetries generated by $\xi^\mu$ and $\epsilon $ lead to Noether charges which take the form
\begin{equation}\label{Noether}
    Q[T,Y^A,U]=\frac{1}{16\pi}\int{d^2z\sqrt{\det\ g_{AB}^{(0)}}\Big{[}-Tg_{vv}^{(1)}-Y^Ag_{vA}^{(1)}-4(U+Y^BA_B^{(0)})A_v^{(1)}\Big{]}},
\end{equation}
where $g_{\mu\nu}^{(i)}$ stand for the $i^{\text{th}}$ term in the $\rho$ expansion; namely, $g_{AB}^{(0)}=\Omega_{AB}$, $g_{Av}^{(1)}=\theta _{A}(z,\bar{z})$, $g_{vv}^{(1)}=-2\kappa$. This expression for the charges follows from the Barnich-Brandt formalism \cite{Glenn}. 

Now, having written down the explicit form for the conserved charges associated to the infinite-dimensional gauge symmetries described above, we are in condition to evaluate them for the case in which we are interested: the black hole in a magnetic external field. However, first we have to actually show that Ernst solution (\ref{Ernst})-(\ref{Ernst2}) can be accommodated in the asymptotic form (\ref{Estas})-(\ref{Estas3}) near the horizon. To do so, let us consider the following change of coordinates
\begin{equation}
  v =  t + \int{\frac{dr'}{f(r')}} \ , \ \ \ \rho =  \int_{2M}^{r}{\Lambda(r',\theta)^2 dr'}\, ,
\end{equation}
and then expand in powers of $\rho$. We get
\begin{align*}
    g_{vv} =&\ -\partial_r f(r)_{|r=2M}\, \rho + \mathcal{O}(\rho^2)\\
    g_{v\theta} =&\ -2\partial_\theta\big{(}\log{|\Lambda(r,\theta)|_{|r=2M}}\big{)}\rho + \mathcal{O}(\rho^2)\\
    g_{\theta \theta} =&\ |\Lambda(r,\theta)|_{|r=2M}^2(2M)^2+\partial_\rho\Big{(}r^2|\Lambda(r,\theta)|^2\Big{)}_{|\rho=0}\rho+\mathcal{O}(\rho^2)\\
    g_{\varphi \varphi} =&\ |\Lambda(r,\theta)|_{|r=2M}^{-2}(2M)^2\sin^2{\theta}+\partial_\rho\Big{(}r^2|\Lambda(r, \theta)|^{-2}\Big{)}_{ |\rho=0}\sin^2{\theta}\hspace{0.2em}\rho+\mathcal{O}(\rho^2)\\
    g_{v\rho} =&\ 1\, ,
\end{align*}
where, in particular, we see that $\kappa = \frac 12 \partial_ r f(r)_{|r=r_+}$; namely, the surface gravity is independent of $B_0$. From this, we can actually evaluate the charges (\ref{Noether}). As $g_{v\phi}^{(1)}$ vanishes and $g_{v\theta}^{(1)}$ happens to be an even function of $\theta $ in the range $[0,\pi]$, both the integrals $Q[Y^A=1]$ and $Q[U=1]$ vanish. The charge $Q[T=1]$, in contrast, gives a non-vanishing result; namely $Q[T=1]={M}/{2}$, which can be written as
\begin{equation}
    Q[T=1]= \frac{\kappa}{2\pi }\, \frac{4\pi r_+^2}{4}.
\end{equation}
We notice from this that the Noether charge associated to the zero-mode ($T=\text{const.}$) of the supertranslation $\xi =T(\theta, \phi)\partial_v$ corresponds to the Wald entropy; more precisely, it gives $Q[T=1]=T_{\text{H}}S_{\text{BH}}$, where $T_{H}=\kappa/(2\pi )$ is the Hawking temperature of the black hole, and $S_{\text{BH}}=\mathcal{A}/4$ is the Bekenstein-Hawking entropy that fulfills the area law (here, $\hbar = k_B = 1$). This result generalizes those obtained in \cite{DGGP1, DGGP2, Puhm} where stationary black holes in asymptotically flat and asymptotically Anti-de Sitter spacetimes were considered using a similar method. Here we have shown that an analogous calculation based on the near horizon expansion works for black hole solutions in presence of backreacting external magnetic fields. 

The computation of Noether charges at the horizon, being those charges associated to near-horizon asymptotic conditions, raises the question as to whether it makes sense to consider the horizon surface as an actual boundary. In fact, while black hole event horizons can well be taken as a boundary of certain spacetime region, it is also true that they are not strictly a boundary to all dynamical observers. Still, one knows that somehow horizon charges do make sense; after all, Wald entropy is an example of them. Understanding this issue requires to be aware of the fact that the physical meaning of a horizon charge -- say the Wald entropy -- has to be referred to special observers -- resp. external observers--. Besides, it is usually the case that a kind of Gauss phenomenon take place and one can also compute other conserved charges from the horizon perspective that happen to agree with the charges computed at infinity with, for instance, the ADM type methods. Such is the case of the angular momentum of stationary black holes, which, as shown in \cite{DGGP1}, is associated to the zero-mode of horizon superrotations. Here, in particular, we are interested in charged black hole solutions immersed in an external magnetic field, which, as mentioned before, acquire a non-vanishing dragging due to the interplay between the electric charge and the external field, and so they turn out to be stationary. We will see below that this leads to a new phenomenon from the horizon perspective, as the non-vanishing Poynting density provides a non-zero (super)rotation charge even in absence of intrinsic spin of the black hole. The superrotation charges computation for these non-static solutions will yield consistent results.

\section{Charged, stationary black holes}

Let us consider charged, stationary black holes in an external magnetic field. By applying the Harrison method with the Reissner-Nordstr\"om solution as a seed, Ernst was also able to obtain an exact solution to Einstein-Maxwell equations that describe such a black hole in Melvin universe\footnote{While uncharged, stationary Ernst solution is asymptotically Melvin, the stationary version of such black hole is not strictly asymptotically Melvin as the ergoregion extends to infinity; see \cite{Gibbons}.} \cite{Ernst}. The spacetime metric reads
\begin{equation}\label{complicada}
    ds^2= |\Lambda (r,\theta )|^2\Big(-f(r)dt^2+\frac{dr^2}{f(r)}+r^2 d\theta^2\Big)+|\Lambda (r,\theta )|^{-2}r^2 \sin^2{\theta}\, (d\phi-\omega (r,\theta ) dt)^2
\end{equation}
where
\begin{align*}
\Lambda (r,\theta ) =&\, 1+\frac{1}{4}B_0^2(r^2 \sin^2{\theta}+q^2\cos^2{\theta})-iB_0 q\cos{\theta}\, , \\ 
f(r)=&\, 1-\frac{2M}{r}+\frac{q^2}{r^2}\, \\
    \omega (r,\theta )=&\, B_0q\bigg{[}-\Big{(}\frac{2}{r}-\frac{2}{r_+}\Big{)}+\frac{B_0^2}{2}\Big(r-r_{+}+rf(r)\cos^2{\theta}\Big)\bigg{]}+\omega_0\, ,
\end{align*}
where $\omega_0$ is a constant that can be absorbed by a local boost, and where $r_+=M+\sqrt{M^2-q^2}$ is the external horizon of the charged black hole. This solution reduces to the Reissner-Nordstr\"om solution with mass $M$ and electric charge $q$ in the case $B_0=0$, while it agrees with the Ernst solution (\ref{Ernst}) in the uncharged case $q=0$. Notice that the function $\Lambda (r,\theta )$ now is non-real, with its imaginary part gathering the interplay between the electric and magnetic components of the solution; $\Lambda (r,\theta )$ becomes real when the product $qB_0$ vanishes. The non-vanishing components of the electromagnetic field are given by
\begin{align*}
        A_\phi(r,\theta) = & \frac{1}{B_0}\bigg{[}1+\Big{(}\frac{\text{Re}(\Lambda(r,\theta))}{|\Lambda(r,\theta)|}\Big{)}\Big{(}\frac{\text{Re}(\Lambda(r,\theta))-2}{|\Lambda(r,\theta)|}\Big{)}\bigg{]},\\
        A_t(r,\theta) = & \frac{2q}{r} + \frac{3\omega(r,\theta)}{2B_0} - A_\phi(r,\theta) \omega(r,\theta),
\end{align*}
where $\text{Re}(\Lambda )$ stands for the real part of $\Lambda $, so that the associated electric and magnetic fields are
\begin{align*}
    H_r+iE_r =&\ \Lambda^{-2} (r , \theta )\bigg{[} i\frac{q}{r^2}(2-\text{Re}(\Lambda (r, \theta ))\, )+B_0(1-\frac{1}{2}iB_0q\cos{\theta})\Big{(}1-\frac{q^2}{r^2}\Big{)}\cos{\theta} \bigg{]},\\
    H_\theta + i E_\theta = &\ -B_0\Lambda^{-2}(r, \theta )\, (1-\frac{1}{2}iB_0q\cos{\theta})f^{1/2}(r)\sin{\theta}.
\end{align*}
This way of writing the electric and magnetic components of the field strength as the real and imaginary parts of complex functions is standard in the solution generating methods employed in \cite{Ernst}, cf. \cite{ErnstWild}. 

We see in the spacetime metric above that an off-diagonal term $g_{\phi t}$ appears as a result of the interplay between the electric and the magnetic fields. Such term represents a differential rotation of the spacetime, which is controlled by the function $\omega (r,\theta )$. The latter obviously vanishes when either $B_0$ or $q$ is zero. However, this is not the only new effect relative to (\ref{Ernst})-(\ref{Ernst2}) that appears when $qB_0\neq 0$. Another interesting feature is a conical singularity that would appear at the poles $\theta = 0 , \, \pi $ unless one demands $\phi $ to take a specific periodicity condition that depends on $q$ and $B_0$. To see this explicitly, let us consider the induced metric on the constant-$t$, constant-$r$ 2-dimensional space; namely
\begin{equation*}
    ds^2_{2}=|\Lambda (r,\theta )|^2\bigg{(}r^2d\theta^2+\frac{r^2\sin^2{\theta}d\phi^2}{|\Lambda (r,\theta )|^2}\bigg{)}\, ,
\end{equation*}
from what we see that the constant-$\theta$ circumferences take the values ${2\pi {|\Lambda (r,\theta )|^{-2}} r\sin{\theta}}$. Since the function $|\Lambda (r,\theta )|$ does not depend on $r$ at the poles, namely $\partial_r \Lambda (r,\theta )_{|\theta=\frac{\pi}{2}\pm \frac{\pi}{2}}=0$, and $|\Lambda(r,\theta =0)|=|\Lambda(r,\theta =\pi)|$, then we can define the constant $|\Lambda_0|\equiv |\Lambda(r,\theta =0)|$ and demand $\phi $ to be periodic and to take values in the range $\phi \in [0,2\pi |\Lambda_0|^2]$; notice that $|\Lambda_0|=1$ when $qB_0=0$. With this, the solution becomes smooth at the poles. Notice that it is important to take into account the $q$-dependent periodicity of the angular coordinate $\phi $ when computing the conserved charges, as such computation involves integrating on constant-$v$, constant-$r$ surfaces. This detail will be important in the calculation below.

In order to compute the horizon charges, we may first have to write the geometry (\ref{complicada}) in its near horizon form (\ref{Estas})-(\ref{Estas3}). To do so, we can consider a change of coordinates of the form 
\begin{align*}
    dv =&\ dt+\frac{dr}{f(r)},\\
    d\varphi =&\ d\phi + \frac{\omega(r,\theta)}{f(r)}dr+h(r,\theta)d\theta,\\
    d\rho =&\ |\Lambda(r,\theta)|^2 dr + g(r,\theta)d\theta,
\end{align*}
with $h(r,\theta)$ and $g(r,\theta)$ being two specific functions that vanish at the horizon $r_+$; namely
\begin{align*}
    h(r,\theta) =&\ \int_{r_+}^r{{f^{-1}(r')}\partial_{\theta }\omega(r',\theta)dr'},\\
    g(r, \theta) =&\ \int_{r_+}^r{{\partial}_{\theta}|\Lambda(r',\theta)|^2 dr'}.
\end{align*}
Integrating this conditions, we get
\begin{align*}
  v = &\ t + \int{\frac{dr'}{f(r')}}, \\
  \varphi = &\ \phi + \int_{r_+}^{r}{\frac{\omega(r',\theta)}{f(r')}dr'}, \\
  \rho = &\ \int_{r_+}^{r}{|\Lambda(r',\theta)|^2 dr'}.
\end{align*}
With this, metric (\ref{complicada}) takes the near horizon form
\begin{align*}
    g_{vv} =&\ -\partial_{r}f(r)_{|r=r_+}\rho + \mathcal{O}(\rho^2),\\
    g_{v\theta} =&\ -2\partial_\theta\big{(}\log{|\Lambda (r,\theta)|}\big{)}_{|r=r_+}\rho + \mathcal{O}(\rho^2),\\
    g_{v\varphi} =&\ -|\Lambda_+(\theta)|^{-4}r_+^2 \sin^2{\theta}\, \partial_\theta\omega (r,\theta )_{|r=r_+}\rho + \mathcal{O}(\rho^2),\\
    g_{\theta \theta} =&\ |\Lambda_+(\theta)|^2r_+^2+\partial_\rho\Big{(}|\Lambda(r, \theta)|^2r^2\Big{)}_{|\rho=0}\rho+\mathcal{O}(\rho^2),\\
    g_{\varphi \varphi} =&\ |\Lambda_+(\theta)|^{-2}r_+^2\sin^2{\theta}+\partial_\rho\Big{(}|\Lambda(r, \theta)|^{-2}r^2\Big{)}_{|\rho=0}\sin^2{\theta}\,\rho+\mathcal{O}(\rho^2),\\
    g_{\theta \varphi} =&\ |\Lambda_+(\theta)|^{-4}r_+qB_0^3\sin^3{\theta}\cos{\theta}\, \rho+\mathcal{O}(\rho^2),\\
    g_{v\rho} =&\ 1\, ,
\end{align*}
where $\Lambda_+ (\theta )\equiv \Lambda(r=r_+,\theta)$; the other components vanish, namely $g_{\rho\rho }=g_{\rho A }=0$. The transformed expression for the electromagnetic field reads
\begin{equation}
    {A} = A_t \ \tilde{d}v - \frac{1}{f(r)}(A_t+\omega(r,\theta)A_\phi)\, \tilde{d}r+A_\phi \, \tilde{d}\varphi-A_\phi\, h(r,\theta)\, \tilde{d}\theta \, .
\end{equation}
In addition, it is also necessary to ensure the gauge fixing condition $A_\rho = 0$, which is achieved by performing a gauge transformation
\begin{equation}
    {A}\, \to \, \Tilde{A} = {A} + \Tilde{d}\zeta\, , \ \ \text{with}\ \  \  \zeta(\theta,r) = \int_{r_+}^{r}{{f^{-1}(r')}(A_t+\omega(r',\theta)A_\phi)dr'}\, .
\end{equation}
This yields $\partial_\theta\zeta = \frac{3}{8}B_0^2q(r^2-r_+^2)\sin{2\theta}$, and finally the electromagnetic field takes the form
\begin{equation}
    \Tilde{A} = A_t \Tilde{d}v + A_\phi \Tilde{d}\varphi - \Big{[}A_\phi h(r,\theta)-\partial_\theta \zeta(\theta,r)\Big{]}\Tilde{d}\theta \, ,
\end{equation}
which in the near horizon yields the following components
\begin{align*}
    A_\theta^{(0)} =&\ 0\, , \\
    A_\varphi^{(0)} =&\ A_\phi(r=r_+)\, , \\
    A_v^{(1)} =&\ |\Lambda_+(\theta)|^{-2}\Big{[} \frac{q}{r_+^2} + \frac{3B_0^2q}{4}\Big{(}1+r_+\partial_r f(r)_{|r=r_+}\cos^2{\theta}\Big{)} - \partial_\theta \omega(r,\theta)_{|r=r_+}A_\varphi^{(0)}\Big{]}\, .
\end{align*}

What we have shown here is that the stationary, charged Enrst solution, describing a Reissner-Nordstr\"om in asymptotically Melvin magnetic universe, can be accommodated in the near horizon form of Einstein-Maxwell solutions studied in reference \cite{Puhm}. This implies that the Noether charges (\ref{Noether}) turn out to be finite and integrable for such a spacetime, and they can be explicitly computed from the horizon viewpoint. We do this in the following section. 

\section{Horizon charges for magnetized black holes}

Taking into account the periodicity of $\varphi $ and the determinant ${\det\ g_{AB}^{(0)}}=r_+^4\sin^2{\theta}$, we evaluate the charges (\ref{Noether}) and get the final result
\begin{align}
    Q[T=1]=&\ \frac 12 \Big( 1+ \frac 32 B_0^2q^2 + \frac{1}{16} B_0^4q^4\Big)\, \sqrt{M^2-q^2},\label{La1}\\
    Q[Y^A=1]=&\ \frac{B_0 q^3}{2} \frac{(1-q^2B_0^2-\frac{1}{16}q^4B_0^4 )}{(1+\frac 32 q^2B_0^2 + \frac{1}{16}q^4B_0^4)} \, \delta^{A \varphi } , \label{La2}\\
    Q[U=1]=&\ q\Big{(}1-\frac{B_0^2q^2}{4}\Big{)} . \label{La3}
\end{align}

We see from here that, unlike the solution (\ref{Ernst}), which had only non-vanishing Wald charge $Q[T=1]$, solution (\ref{complicada}) exhibits non-vanishing zero-modes for both supertranslations and superrotations, all of them being $B_0$- and $q$-dependent. In particular, charge (\ref{La1}) gives
\begin{equation}
Q[T=1]=  |\Lambda_0|^2\, \frac{r_+-r_-}{4} = \frac{\kappa}{2\pi } \, \frac{\mathcal{A}}{4} 
\end{equation}
where one has to notice that the factor $|\Lambda_0|^2$ is precisely the one needed to correctly reproduce the horizon area $\mathcal{A}$ due to the $B_0$- and $q$-dependence in the periodicity of $\varphi$. In other words, the dependence on the external magnetic field and on the black hole charge combine in the precise form for the supertranslation charge to reproduce 
\begin{equation}
Q[T=1] \, = \, T_{\text{H}}\, S_{\text{BH}}, 
\end{equation}
now for stationary, charged black holes. This seems consistent with the following result for the mass \cite{Gibbons2}
\begin{equation}
\bar{E}\, =\,  M\,\Big( 1+ \frac 32 B_0^2q^2 + \frac{1}{16} B_0^4q^4\Big)\, ,\label{MasaGi}
\end{equation}
which follows from the Komar integral. In \cite{Gibbons2}, the authors discuss (\ref{MasaGi}) in the context of black hole mechanics, considering different thermodynamical ensembles, including the one in which $B_0$ is permitted to vary adiabatically. However, the impossibility of deriving the energy expression (\ref{MasaGi}) as an integrable conserved charge was observed in \cite{AstorinoCompere}, where the following result for the mass was obtained
\begin{equation}
{E}\, =\,  M\,\Big( 1+ \frac 32  B_0^2q^2 - \frac {1}{M^2}  B_0^2q^4 + \frac{1}{16} B_0^4q^4\Big)^{1/2}\, ,\label{MAstorino}
\end{equation}
yielding $\bar{E}-{E}=\mathcal{O}(MB_0q^2,B_0q^4/M)$. Expression (\ref{MAstorino}) turns out to be integrable, it reduces to the mass of vacuum Einstein equations when $B_0=0$, and it fulfills the Christodoulou-Ruffini relation between the black hole mass and the entropy. In fact, result (\ref{MAstorino}) is totally consistent with black hole thermodynamics, leading to sensible chemical potentials. Following from an integrable expression of a covariant method, expression (\ref{MAstorino}) seems to be the correct result of the Enrst-Wild black hole mass. Since in our near-horizon approach we directly have access to the value $T_{\text{H}}S_{\text{BH}}$, rather than the mass, and given that the values for $T_{\text{H}}$ and $S_{\text{BH}}$ are consistent with those in the literature, our computation turns out to be consistent with (\ref{MAstorino}).

Probably the most remarkable result is the superrotation charge (\ref{La2}), which is non-zero despite the black hole not having intrinsic spin. As anticipated, this non-vanishing result is due to the Poynting density that solution (\ref{complicada}) exhibits. This induces a dragging effect in the spacetime, breaking staticity. The result (\ref{La2}) agrees with the one computed in the literature \cite{Gibbons}, here having computed it from the near horizon perspective, showing the Gauss phenomenon to hold for this type of solution. The superrotation charge can also be written as follows
\begin{equation}\label{coso}
Q[Y^\theta=1]\, =\, 0\ \ , \ \ \  \  Q[Y^{\varphi }=1]\, =\,  |\Lambda_0|^{-2}\frac{B_0 q^3}{2} (1-q^2B_0^2-\frac{1}{16}q^4B_0^4 ) \, ,
\end{equation}
Adding intrinsic spin to the solution \cite{ErnstWild} would make the superrotation charge $Q[Y^{\varphi }]$ to acquire extra dependence on the Kerr parameter $a=J/M$. Kerr black holes were studied from the near horizon perspective in \cite{DGGP1}. Notice, however, that the rotation charge $Q[Y^{\varphi }=1] $ in (\ref{coso}) is associated to the vector $\xi = Y^{\varphi }\partial_{\varphi }= \partial_{\varphi } $, with $\varphi $ being the angular coordinate with periodicity $2\pi|\Lambda_0 |^2$, related to the $qB_0$-dependent angular deficit at the poles. Therefore, the charge that represents the angular momentum would rather be the one defined with respect to the vector $\xi =\partial_{\varphi /|\Lambda_0|^2}$, namely $Q[Y^{\varphi }=|\Lambda_0 |^2]$, which amounts to multiply (\ref{La2}) by a factor $|\Lambda_0 |^2$. This yields the simpler expression $Q[Y^{\varphi }=|\Lambda_0 |^2]= \frac 12 B_0 q^3 (1-B_0^2q^2-\frac{1}{16}B_0^4 q^4)$. Next, we have to be aware of the fact that the result for the angular momentum $J$ is not invariant under a residual gauge symmetry that the solution exhibits. For example, there exists an ambiguity in the definition of $A_{\varphi}^{(0)}$; this is explained in detail, for instance, in section V of \cite{Gibbons2} and in references thereof. Therefore, in order to compare the result of $J$ with the one obtained in the literature, we have to be sure we are considering the same gauge. If, with the authors of \cite{Gibbons2}, we choose a gauge $A \rightarrow \tilde{A}=A + c \, \tilde{d}\varphi$ with $c(q,B_0)$ a constant given by $ c(p,B_0) = -A_{\phi}(r_+,\theta = 0) = -A_{\phi}(r_+,\theta = \pi)$, which makes the angular components of $A_{\mu}$ to vanish at the poles, the angular charge is  
\begin{equation}\label{coso}
\tilde{Q}[Y^{\varphi}  = | \Lambda_0 |^2] \, = \, -q^3B_0\,  \text{Re} ( \Lambda_0 )\, ;
\end{equation}
that is,
\begin{equation}\label{coso}
J\, =  \, -q^3B_0 \Big( 1+  \frac 14 q^2B_0^2 \Big) \, ,
\end{equation}
which agrees with the result obtained in the literature by other methods, cf. \cite{AstorinoCompere}. Together with the other charges computed above, it is consistent with the Smarr formula \cite{AstorinoCompere} and with the first principle of black hole mechanics \cite{Gibbons2}. The value of $J=\tilde{Q}[Y^{\varphi} ] $, of course, flips its sign under $qB_0 \to -qB_0$. It is remarkable that the rotation charge exhibits a non-vanishing angular momentum despite the black hole has no intrinsic spin parameter ($a=0$). This is due to the interplay between the external magnetic field and the black hole electric charge, and it is consistent with what we know about charged black holes in front of a magnetic monopole probe. When adiabatically approaching a monopole to a Reissner-Nordstr\"om black hole \cite{Ray}, the latter starts to acquire angular momentum \cite{HT}. The result above can be seen as a realization of such phenomenon in a backreacting scenario. 

Last, charge (\ref{La3}) corresponds to the effective electric charge at the horizon; consistently, it reduces to $Q[U=1]\simeq q $ in the small $B_0$ limits. Again, this result, obtained here from the near horizon computation, agrees with the results obtained by other methods, cf. \cite{Gibbons}. It exhibits very interesting features, the most salient one being the critical value of the magnetic field, $B_0=\pm 2q^{-1}$, such that the charge vanishes despite $q\neq 0$. Moreover, $Q[U=1]$ can even change its sign relative to $q$ and acquire an arbitrary large absolute value, provided $B_0$ increases sufficiently. The implications of this phenomenon are interesting. For example, we can consider the so-called black hole Meissner effect, namely the expulsion of the external magnetic field by a black hole when extremality is approached \cite{Bicak1, Meissner}. This is reminiscent of the effect exhibited in superconductors, therefore the name. We can verify that the results obtained above for the conserved charges can be considered to study the black hole Meissner effect in a working example\footnote{The simplest way to see the Meissner effect would be resorting to the working example discussed in section VII B of \cite{Gibbons2}: First, one asks the effective electric charge $Q[U=1]$ to vanish (that is, tuning $B_0=2q^{-1}$) and, then, one demands the extremality condition $Q[T=1]=0$ (namely $M=\pm q$). After doing this, one finds that the flux of the magnetic field, $\int_{\Sigma_2 } B = 0$, through the horizon $\mathcal{H}^+=\Sigma_2\times \mathbb{R}_v$ vanishes. This realizes the black hole Meissner effect.} discussed in \cite{Gibbons2}. In relation to the Meissner effect, it is worth mentioning reference \cite{Astorino}, which to the best of our knowledge is the first article in which the map between the magnetized and the non-magnetized black hole at zero temperature is studied.

\section{Gravitating monopoles in external magnetic field}

As a last example, let us consider a magnetically charged black hole in the external magnetic field. This is described by the metric
\begin{equation}\label{melvinR-Nmag}
    ds^{2} = \Lambda^{2}(r, \theta)\, \left( - f(r) dt^{2} + \frac{dr^{2}}{f(r)} + r^{2} d\theta^{2} \right) + \frac{r^{2} \sin^{2}(\theta)}{\Lambda^{2}(r, \theta)} \, d\phi^{2}
\end{equation}
with
\begin{equation}
    \Lambda (r, \theta) = 1 + \frac{B_0^{2}}{4} \left(r^{2} \sin^{2}\theta + p^{2} \cos^{2}\theta \right) - B_0p \, \cos\theta \ , \ \ \ f(r) = 1 - \frac{2M}{r} + \frac{p^{2}}{r^{2}}
\end{equation}
where $p$ is the magnetic charge of the black hole. The angular coordinate $\phi $ is periodic and now takes values in the range $[-C\pi,C\pi]$ with $C$ a constant. In the limits $B_0=0$ and $p=0$ this metric reduces to that of the Reissner-Nordstr\"om and the uncharged Ernst black hole, respectively. The horizons are located at $r_{\pm }= M\pm \sqrt{M^2-q^2}$, and the solution is singular at $r=0$. The electromagnetic potential in this case is given by
\begin{equation}
    A = \frac{-p \, \cos\theta + \frac{B_0}{2}(r^2 \sin^2\theta + p^2 \cos^2\theta )}{\Lambda (r,\theta )} \, d\phi
\end{equation}
which tends to the Dirac monopole solution when $B_0$ is sufficiently small. 

Parameter $C$, which defines the periodicity in $\phi$, is determined by requiring the solution to be smooth at one of the poles, say at $\theta = 0$. This amounts to choose $C=\Lambda^2(\theta=0)=1+B_0^2p(p-B_0)/4$. -- If one instead decided to make the solution smooth at $\theta = \pi $ then $C$ should have taken the value $C=\Lambda^2(\theta=\pi )=1+B_0^2p(p+B_0)/4$ --. It is impossible to make both poles $\theta = 0$ and $\theta = \pi$ smooth at the same time, and this has a clear physical interpretation: Representing a magnetic charge in an external magnetic field, solution (\ref{melvinR-Nmag}) needs the presence of a cosmic string that provides the tension for keeping the black hole at rest. Such cosmic string is precisely the conical defect pinching the horizon at one of the poles and extending to infinity. Notice that this is consistent with the fact that $\Lambda^2 (\theta = 0)$ and $\Lambda^2 (\theta = \pi )$ get interchanged under $p,B_0 \to -p,B_0$ or $p,B_0 \to p,-B_0$, while both $\Lambda^2 (\theta = \pi/2\pm \pi/2 )$ remain unchanged under $p,B_0 \to -p,-B_0$. This string is similar to the one appearing in the $C$-metric, with the difference being that here the string provides the black hole with the tension that cancels the acceleration that otherwise it would take due to the magnetic force imposed by the external field. By performing the near horizon analysis on metric (\ref{melvinR-Nmag}), one finds that also in this magnetically charged case the zero-mode of the supertranslation charge reproduces the Wald entropy formula $Q[T=1]=T_{\text{H}}S_{\text{BH}}$, once one takes into account the correct $\phi$-periodicity. The near horizon analysis of the $C$-metric type solutions have recently been performed in \cite{nosotros}. 

\[\]

The authors thank Andr\'es Anabal\'on, Laura Donnay, Hern\'an Gonz\'alez, and Julio Oliva for discussions on related subjects. This work has been supported by CONICET and ANPCyT through grants PIP-1109-2017, PICT-2019-00303.

  \end{document}